\documentclass[epj]{svjour}
\usepackage{pifont}
\usepackage[dvips]{graphicx}
\usepackage{array}
\usepackage{amssymb}
\usepackage{amsmath}
\usepackage{hhline}
\usepackage{longtable}
\begin{document}
\title{Propagation of travelling waves in sub-excitable systems driven by noise and periodic forcing}
\author{Fen-Ni Si\inst{1,2} \and Quan-Xing Liu\inst{3} \and Jin-Zhong Zhang\inst{2,4} \and Lu-Qun Zhou\inst{2}}                     
\offprints{sifenni@163.com (Fen-Ni Si)}          

\institute{China Academy of Engineering Physics, Mianyang, Sichuan
621000, People's Republic of China \and Department of Physics,
Peking University, Beijing 100871, People's Republic of China \and
Department of Mathematics, North University of China, Taiyuan,
Shan'xi 030051, People's Republic of China \and Department of Physics, Northeastern University, Boston, MA 02115, USA}
\date{Received: date / Revised version: date}

\abstract{It has been reported that traveling waves propagate
periodically and stably in sub-excitable systems driven by noise
[Phys. Rev. Lett. \textbf{88}, 138301 (2002)]. As a further
investigation, here we observe different types of traveling waves
under different noises and periodic forces, using a simplified
Oregonator model. Depending on different noises and periodic forces,
we have observed different types of wave propagation (or their
disappearance). Moreover, the reversal phenomena are observed in
this system based on the numerical experiments in the
one-dimensional space. As an explanation, we regard it as the effect
of periodic forces. Thus, we give qualitative explanations to how
reversal phenomena stably appear, which seem to arise from the
mixing function of the periodic force and the noise. And the output
period and three velocities (the normal, the positive and the
negative) of the travelling waves are defined and their relationship
with the periodic forces, along with the types of waves, are also
studied in sub-excitable system under a fixed noise intensity.
} 

\maketitle

\section{Introduction}

The effects of noise on nonlinear systems are the subject of intense
experimental and theoretical investigations. Noise can induce
transition~\cite{Horsthemke,RM2006},
bifurcations~\cite{1742-5468-2007-07-P07016}, and stochastic
resonance~\cite{Gammaitoni,Hz1998,KW1995,PhysRevLett.74.2130}.
Especially, in Ref.~\cite{PhysRevLett.74.2130} the synchronization
of spatiotemporal patterns were observed in an excitable medium by
the numerical evidence. Moreover noise can enhance propagation in
arrays of coupled bistable
oscillators~\cite{Lindner,PhysRevLett.76.2609,PhysRevE.54.3479,PhysRevA.38.983}.
In an excitable system, an external periodic forcing can
dramatically change its behavior. As reported by previous documents,
the phase locking, quasi-periodicity, period doubling, and chaos
were observed~\cite{Feingold}. The temporal evolution of the
concentration patterns has been modeled by partial differential
reaction-diffusion equations. Such models include oscillatory,
excitable or bistable systems with either none, one or two linearly
stable homogeneous states~\cite{meron,Mikhailov}. It is also well
known that in sub-excitable systems noise also can induce travelling
waves~\cite{Kadar}, drive avalanche behavior~\cite{Wang}, and
sustain pulsating patterns and global
oscillations~\cite{PhysRevLett.82.3713}. Sub-excitable systems under
noises and periodic forcing are able to send out travelling and
spiral waves. Belousov-Zhabotinsky (BZ) reaction~\cite{Zakin,Field}
is a popular symbol in nonlinear dynamical realm to study excitable
and sub-excitable system. It has been widely agreed that the noise
and periodic forcing play a very important role in wave propagation
and stability.

Recently, it was observed that noise can support wave propagations
in sub-excitable~\cite{Zhou2002,PhysRevLett.82.3713,Kadar} systems
due to a noise-induced
transition~\cite{PhysRevLett.87.078302,Garcia-Ojalvo}. In these
studied, the medias are static, and transports are governed by
diffusion~\cite{1742-5468-2007-07-P07013}. However in many
situations, the medias are not static but subject to a motion. For
example, stirred by a flow, or by the periodic forcing, the
convective-like phenomena were observed due to electric field in
Ref.~\cite{PhysRevLett.76.3854}, which occurs especially in chemical
reactions in a fluid environment. In such cases,  diffusive
transport usually dominates only at small spatial scales while
mixing due to the flow in much faster at large scale. In
Refs.~\cite{PhysRevLett.91.084101,PhysRevE.66.066208,1367-2630-7-1-018},
the authors shown that in an inhomogeneous self-sustained
oscillatory media, an increasing rate of mixing can lead to a
transition to a global synchronization of the whole media.
Especially, Ref.~\cite{PhysRevLett.91.150601} showed that the
interplay among excitability, noise, diffusion and mixing can
generate various pattern formation in a 2D FigzHugh-Nagumo (FHN)
model subject to the advection by a chaotic flow. Here, we research
the effect of noise and periodic forcing on sub-excitable systems
using the Oregonator model in one dimension, which advances from the
BZ reaction. The reversal phenomenon is not observed in the previous
documents about the propagation of travelling waves. It is founded
in this paper, which relates to a new concept. In
Ref.~\cite{Albanese}, Richard A. Albanese refers to reversal concept
in wave propagation concern extraction of information about distant
structural features from the measurements of scattered waves, but it
is irrelevant the excitable system.

In our paper, we define that the general traveling waves propagate
forward in one constant direction and vice versa. Under this
definition, we found the reversal phenomena in our numerical
simulations. However in our simulation we have discovered that after
some time, the waves change its propagation direction and turn
backward to travel in the opposite direction. The traveling waves
propagate forward and backward alternately and periodically. That is
called the reversal phenomenon (see the supplementary material
on-line movie for this phenomenon, \emph{Movie-0}). In the
mathematical language, the definition of definition of an ``reversal
phenomenon" as at time $t$, the spatial position of a traveling wave
front is at point $x$. After some time $\Delta t>0$, the wave front
reaches point $x$ again with reversal direction. In fact, this
phenomenon was observed in the FHN model by numerical
simulations~\cite{PhysRevLett.91.150601}. In this article, our
focuses are the effect of the periodic forcing and noise on the
propagation of traveling waves in sub-excitable systems.
\section{Model}

Most of the systems we are interested in reside in a d-dimensional
world. This means that our variables (fields or concentrations)
depend on time and space. In present paper, the starting
deterministic model~\cite{Zhou2002} is based on partial differential
equations, and when the randomness is introduced we transform them
into stochastic partial differential equation. A representative
example is the deterministic reaction-diffusion equation,
\begin{equation}\label{eq:0}
  \frac{\partial \phi({\bf x},t)}{\partial t}=f(\phi({\bf
  x},t),\mu)+\mathcal{D}\nabla^2 \phi({\bf x},t),
\end{equation}
where $\phi({\bf x},t)$ represent the density of a physical
observable, $f(\phi({\bf x},t),\mu)$ is a nonlinear function of the
field $\phi$ and $\mu$ denotes the relevant control parameter.  The
above equation can be made more complicated when considering vector
fields, higher-order derivatives, or nonlocal operators. The effect
of fluctuations is introduced through a stochastic process or noise
$ \xi({\bf x}, t)$ with well controlled statistical properties. As a
result, we expect that the new equation governing our system will
have the generic form
\begin{equation}\label{eq:01}
  \frac{\partial \phi({\bf x},t)}{\partial t}=f(\phi({\bf
  x},t),\mu)+\mathcal{D}\nabla^2 \phi({\bf x},t)+ g(\phi)\xi({\bf
  x},t).
\end{equation}

We take into account this standard example of stochastic partial
differential equation and the two-variable Oregonator
model~\cite{Field,Tyson} that is famous for its convenience to study
the property of diffusion-reaction systems. Our modified model adds
both noise and the periodic forces, which is,
\begin{subequations}\label{eq:1}
\begin{equation}\label{eq:1a}
\frac{\partial u}{\partial
t}=\frac{1}{\varepsilon}f(u,v)+D_{u}\nabla^{2}u+(D_{\rm
s}\xi(t)+E(t))\frac{\partial u}{\partial x},
\end{equation}
\vspace{-0.5cm}
\begin{equation}\label{eq:1b}
  \frac{\partial v}{\partial
t}=g(u,v)+D_{v}\nabla^{2}v+(D_{\rm s}\xi(t)+E(t))\frac{\partial
v}{\partial x},
\end{equation}
\end{subequations}
where $f(u,v)=u(1-u)-fv\frac{u-q}{u+q}$, and $g(u,v)=u-v$.
$\nabla^{2}$ is the Laplacian operator in Cartesian coordinates. $u$
and $v$ represent the concentration of $\rm HBrO_{2}$ and the
catalyst $2 \textrm{Ce}^{4+}$, respectively. Here, the external
electric field can be considered as a spatially uniform electric
field, which includes two parts: stochastic forcing, $\xi(t)$ (we
use the notation of $\xi(t)$ for irrelevant in space) and the
periodic force , $E(t)$. Both of them depend on time $t$ with the
form of Gaussian noise and with the periodic function respectively.
$\xi(t)$ denotes Gaussian white noise with $\langle\xi(t)\rangle=0$
and $\langle\xi(t)\xi(t^{'})\rangle=2D_{\rm s}\delta(t-t')$, a
typical temporally varied Gaussian white noise and  here $D_{\rm s}$
is the intensity of noise. $E(t)$ is the periodic force, where the
sine periodic force is chosen, $E(t)=F\sin(\frac{2\pi}{T_{\rm
in}}t)$. $F$ and $T_{\rm in}$ are the intensity of the periodic
forcing and the input period, respectively. The effect of electric
field is convective-like, as discussed in
Ref.~\cite{PhysRevLett.76.3854}. $D_{u}$ and $D_{v}$ are the
dimensionless diffusion coefficients of $u$ and $v$, $D_{u}=1.0$,
$D_{v}=0.6$~\cite{Jahnke}.

The dynamical system~\eqref{eq:1} is simulated in one-dimensional space by
Euler-Maruyama Method~\cite{DJHigham} with zero flux boundary
conditions on a space length of $500$ elements. The space step is
$\Delta x=0.15$ space unit and the time step is $\Delta t=10^{-3}$
time unit. These parameters are chosen to make the simulation
process relatively stable and the information of wave propagation
can be relatively all-around. In order to avoid the simulation going
to the negative region of $u$, we let $u>q$, if $u<q$, $u=q$, according to the method of
the Refs.~\cite{Skaggs,Winfree}. The simulations have been done as
follows: we excite $3$ elements
 at the left boundary in the our systems under sub-excitable state in the one-dimensional space,
 which serves as the wave
source. The leapfrog method for the advection terms; an implicit
method for the diffusion terms; and a simple explicit Euler method
for the reaction terms. Letting $U_{j}^n\approx u(x_j,t_n)$ and
$V_{j}^n\approx v(x_j,t_n)$, where $x_{j+1}-x_{j}=\Delta x$ and
$t^{n+1}-t^{n}=\Delta t$, results in the following discretised
system:
\begin{eqnarray}\label{eq:method1}
\frac{U_{j}^{n+1}-U_{j}^{n}}{\Delta
t}&=&\frac{1}{\varepsilon}f(U_{j}^{n},V_{j}^{n})+D_{u}\Big[\frac{U_{j+1}^{n+1}-2U_{j}^{n+1}+U_{j-1}^{n+1}}{(\Delta x)^2}\Big]  \nonumber\\
  && +(D_{\rm s}\xi(t)+E(t))\Big[\frac{U_{j+1}^n -U_{j-1}^n}{2 \Delta
  x}\Big],\\
\frac{V_{j}^{n+1}-V_{j}^{n}}{\Delta
t}&=&g(U_{j}^{n},V_{j}^{n})+D_{v}\Big[\frac{V_{j+1}^{n+1}-2V_{j}^{n+1}+V_{j-1}^{n+1}}{(\Delta x)^2}\Big]  \nonumber\\
  && +(D_{\rm s}\xi(t)+E(t))\Big[\frac{V_{j+1}^n -V_{j-1}^n}{2 \Delta
  x}\Big], \nonumber\\
\end{eqnarray}
where $D_{\rm s}$, $F$, and $T_{\rm in}$ are the control parameters.
Note that here the noise term $\xi(t)$ is Gaussian and white. This
is a very reasonable assumption for internal noise, which represents
many irrelevant degrees of freedom evolving very short temporal and
spatial scales. It has a probability density function of the normal
distribution (also known as Gaussian distribution). In other words,
the values that the noise can take on are Gaussian distributed.

\section{Result}

Extensive testing is performed through numerical simulations to the
described model~\eqref{eq:1} and the qualitative results are shown
this section.

\subsection{Sub-excitation and reversal phenomena}
\begin{figure}
  \includegraphics[width=8.5cm]{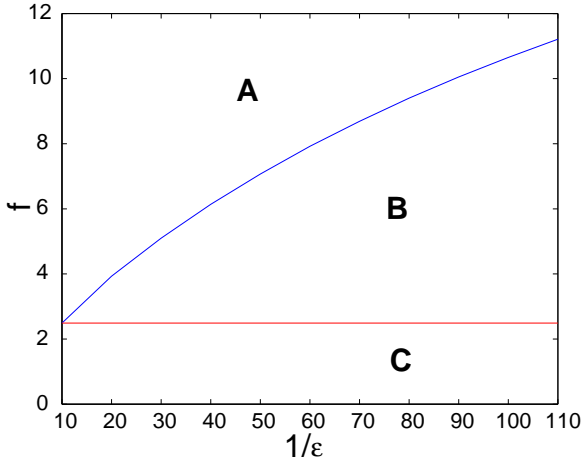}\\
  \caption{(Color online) The phase diagram of the $\varepsilon$ versus $f$.}\label{fig:1}
\end{figure}
$q$, $\epsilon$ and $f$ are parameters related to the BZ kinetics,
determining the sub-excitation of the system. $q=0.002$
\cite{Tyson}. First we simulate the dynamical system~\eqref{eq:1}
without noise and periodic force to confirm the sub-excitable
region. The result is shown in Fig.~\ref{fig:1}. In region (A) no
traveling waves are produced in the system; in region (B) traveling
waves are sent out and propagating in the system; region (C) is not
included in the excitable confine. In the area between region (A)
and (B), traveling waves are sent out but die away gradually, which
is so-called the sub-excitable region, corresponding to the blue
line in Fig.~\ref{fig:1}. Here we set $\epsilon=0.1$ and $f=2.435$
to fix our system into in a sub-excitable region. When the
parameters $\epsilon$ and $f$ are deeply in the sub-excitable
domain, the same results are observed by the numerical experiments,
such as, $\epsilon=0.025$, $f=2.435$ and $\epsilon=0.025$, $f=4.0$.

It turns out that the influence of noise is rather important in the
sub-excitable media. There is an extreme case for the
system~\eqref{eq:1}, that is, when the intensity of noise equals to
$0$, i.e. $D_{\rm s}=0$, and only the periodic force is present, the
different types of wave propagation are summarized in the
Table~\ref{tab:1}. Table~\ref{tab:1} reveals that there exists
critical input period $T_{\rm in}^{*}$ for each intensity of
periodic force $F$ ($F>10.0$), so that the system creates traveling
waves periodically and waves propagate persistently. If $F<10.0$,
the system can only send out one wave, and there is no critical
input period $T_{\rm in}^{*}$; The system sends out traveling waves
periodically but the waves disappear quickly when the $T_{\rm
in}>T_{\rm in}^{*}$ (see \emph{Movie-1} for this case, where $D_{\rm
s}=0$, $T_{\rm in}=8.0$, and $F=10.0$); when $F=10.0$, $T_{\rm
in}^{*}=5.0$; $F=20.0$, $T_{\rm in}^{*}=4.0, 5.0$; $F=30.0$, $T_{\rm
in}^{*}=3.0, 4.0, 5.0$.
\begin{table*}
\caption{\label{tab:1}Traveling waves propagation under different
$F$ and $T_{\rm in}$ with noise intensity $D_{\rm s}=0$; symbol $-1$
indicates the system sends out only one wave; symbol $-$ indicates
the system sends out traveling waves periodically but the waves
disappear quickly; symbol $+$ indicates the system sends out
traveling waves periodically and the waves can propagate
persistently.}
\begin{tabular}{c|ccccccc}\hline
  & $T_{\rm in}<2.0$ & $T_{\rm in}=2.0$ & $T_{\rm in}=3.0$ & $T_{\rm in}=4.3$ & $T_{\rm in}=5.0$ & $T_{\rm in}=6.0$ & $T_{\rm in}>6.0$\\
\hline
$F<10.0$ &   $-1$       &    $-1$      &   $-1$       & $-$           &       $-$   &    $-$       &    $-$       \\
$F=10.0$ &   $-1$      &    $-1$     &   $-1$       & $-1$         &       $+$  &    $-$       &    $-$       \\
$F=20.0$ &   $-1$       &    $-1$      &   $-1$       &  $+$          &       $+$   &    $-$       &    $-$       \\
$F=30.0$ &   $-1$      &    $-1$      &   $+$        &  $+$ & $+$
&    $-$       &    $-$       \\ \hline
\end{tabular}
\end{table*}

\begin{figure}
  \includegraphics[width=8.5cm]{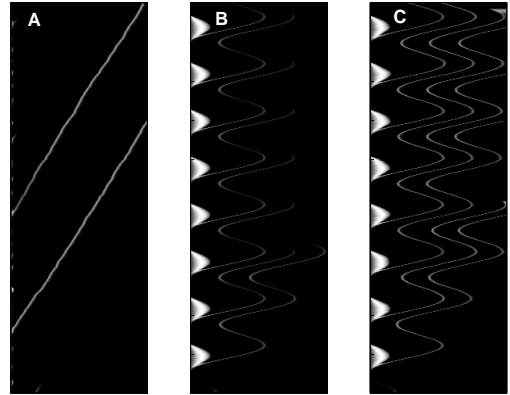}\\
  \caption{(Color online) The spatiotemporal plot of the variable $u$ for the
  system~\eqref{eq:1}, where the ordinate is time evolution and the
   abscissa represents the spatial location. The white part
   indicates wave front. (A) only noise present, $D_{\rm s}=14.0$, and $F=0$
   (see the \emph{Movie-2}, additional movies format available from Journal Web); (B) Both noise and periodic
forcing present, $D_{\rm s}=10.0$, $F=10.0$, and
  $T_{\rm in}=8.0$ (see the \emph{Movie-3}, additional movies format available from Journal Web); (C) Both noise and periodic
forcing present, $D_{\rm s}=14.0$, $F=10.0$, and $T_{\rm in}=8.0$
(see the \emph{Movie-4}, additional movies format available from
Journal Web).}\label{fig:2}
\end{figure}
\begin{figure}
  \includegraphics[width=8.5cm]{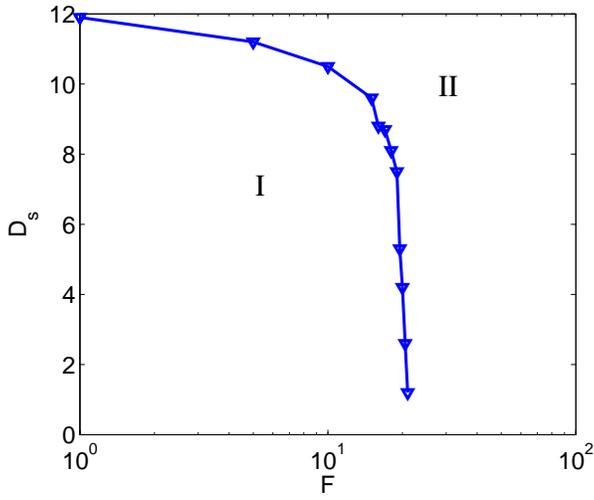}\\
  \caption{(Color online) Phase diagram about the reversal phenomena with respect
  to $D_{\rm s}$--$F$ parameter space. The parameters
  are the same as Fig.~\ref{fig:2}, but $T_{\rm in}=8.0$. The reversal waves will
  die out quickly within region I and propagate persistently within region II.
  Note that the log scale for $T_{\rm in}$. ($\blacktriangledown$, the critical
  values from simulation results.) }\label{figadd1}
\end{figure}

Now, we turn on the noise and the periodic forces in the
system~\eqref{eq:1} and investigate their effects when the intensity
of noise takes different values.
 Fig.~\ref{fig:2} shows that traveling waves propagate with
different noise and periodic forces. The abscissa is the spatial
location and the ordinate is the evolution of time. The white part
indicates wave crests. As shown in Fig.~\ref{fig:2}(A), if only
noise present, traveling waves are produced irregularly but they can
propagate stably [see \emph{Movie-2}]. If periodic force and noise
both present, travelling waves are produced periodically and
reversal phenomena appear, however traveling waves die out quickly
when the intensity of noise is small [see \emph{Movie-3}]. The
travelling waves are produced periodically and they propagate
stably, and reversal phenomena also appear when the intensity of
noise is increased [see \emph{Movie-4}], as shown in
Fig.~\ref{fig:2}(C). In Fig.~\ref{figadd1}, we show the phase
diagram about the reversal phenomena with respect to the $D_{\rm
s}$--$F$ parameter space, in which the reversal travelling waves
will die out quickly within the region I but propagate persistently
within the region II. One can see that the reversal waves
sensitively depend on the intensity of periodic forces and the noise
intensities for the fixing $T_{in}$. For example (see
Fig.~\ref{figadd1}), the emergence of reversal waves is a sensitive
relationship at before and after the certain critical value. For the
large $F$, the curve is sharp decreasing, otherwise the cure is slow
decreasing. So we can obtain conclusion from Figs.~\ref{fig:2} and
\ref{figadd1}, that the stabilization of traveling waves are due to
noise, but the periodicity owe to the function of periodic force.
Periodical and stable traveling waves, as well as reversal
phenomena, are produced in sub-excitable system driven by noise and
periodic force.

\begin{figure}
  \includegraphics[width=8.5cm]{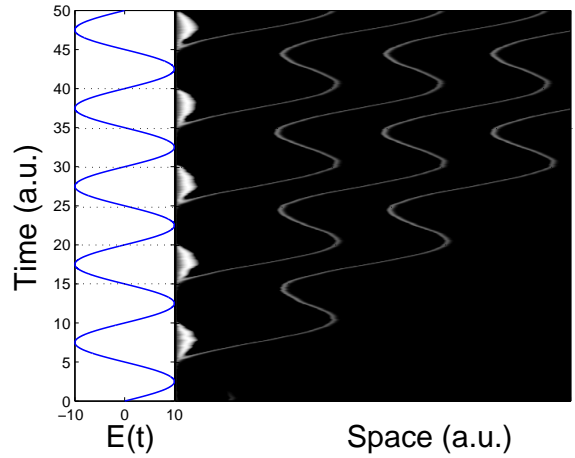}\\
  \caption{(Color online) The propagation of traveling waves with the periodic force
  $E(t)=F\sin(\frac{2\pi}{T_{\rm in}}t)$.
  The ordinate is time evolution. For the left window the
   abscissa represents $E(t)$ and for the right the
   abscissa is space location. The left and right window share the same ordinate which is
time. The parameters are $D_{\rm s}=14.0$, $F=10.0$, and $T_{\rm
in}=8.0$.}\label{fig:3}
\end{figure}
\begin{figure}
  \includegraphics[width=8.5cm]{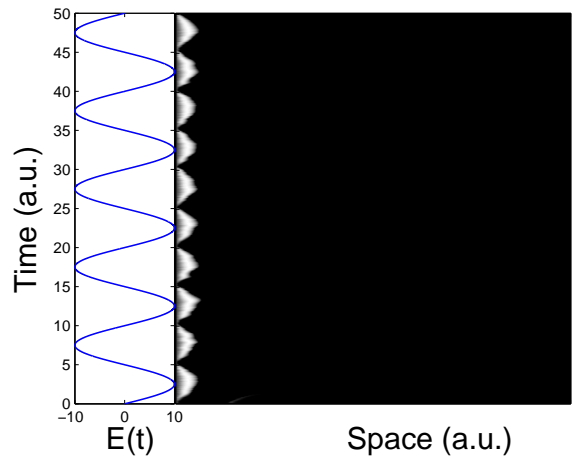}\\
  \caption{(Color online) The propagation of travelling waves with the periodic
  forcing $E(t)=E_{-}(t)$ (or $E(t)=E_{+}(t)$). The ordinate is time evolution. For the left window the
   abscissa represents $E(t)$ and for the right the
   abscissa is space location.                          The left and
right window share the same ordinate which is time. The parameters
are $D_{\rm s}=14.0$, $F=10.0$, and $T_{\rm in}=8.0$.}\label{fig:4}
\end{figure}

To characterize the relation between the periodic force and reversal
phenomena, we give some qualitative explanations. Fig.~\ref{fig:3}
shows the relationship between the propagation of traveling waves
with the periodic force $E(t)=F\sin(\frac{2\pi}{T_{\rm in}}t)$. The
left window of Fig.~\ref{fig:3} is the periodic force whose abscissa
is $E(t)$ and the right window represents the spatiotemporal plot of
variable $u$ whose abscissa is the spatial location. The left and
right window share the same ordinate which is time. From
Fig.~\ref{fig:3} we can easily obtain that corresponding to each
input period the system sends out a traveling wave, which is $1:1$
frequency locking. The dotted line in the left window of
Fig.~\ref{fig:3} divide out the positive and negative parts of the
periodic force. We find out that if the periodic force is negative,
traveling waves propagate forward, whereas if the periodic force is
positive, traveling waves propagate backward in the opposite
direction. Thus reversal phenomena appear.

For further investigation, we test several different kinds of
periodic force for system~\eqref{eq:1}. They are
$E_{\sqcap}(t)=(-1)^{\lfloor \frac{2t}{T_{\rm in}}\rfloor}F$ (the
rectangle periodic force, here the $\lfloor n\rfloor$ denotes the
integer of the $n$), $E_{-}(t)=-|F\sin(\frac{2\pi}{T_{\rm in}} t)|$
and $E_{+}(t)=|F\sin(\frac{2\pi}{T_{\rm in}} t)|$. When the periodic
forcing is $E_{\sqcap}(t)$, travelling waves are produced
periodically and they propagate stably, and reversal phenomena
appear (resemble the Fig.~\ref{fig:2}(C)). When the periodic force
is $E_{-}(t)$ or $E_{+}(t)$, there are only foundations formed but
no travelling waves are sent out, as shown in Fig.~\ref{fig:4}. So
the reversal phenomenon is due to the alternation of the positive
and negative values of the periodic force. If each value of the
periodic force was positive (or negative), no travelling waves will
be sent out and of course no reversal phenomena will appear.

\subsection{The output period and the three velocities}

\begin{figure}
  \includegraphics[width=8.5cm]{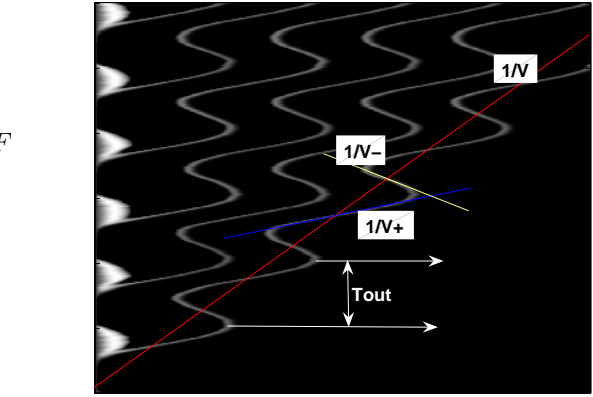}\\
  \caption{(Color online) The sketch map of $T_{\rm out}$, $V$, $V_{+}$,
  and $V_{-}$ with $D_{\rm s}=14.0, F=10.0, T_{\rm in}=8.0$.}\label{fig:5}
\end{figure}

In this section we focus on the effect of the periodic force
$E(t)=F\sin(\frac{2\pi}{T_{\rm in}}t)$ on the propagation of
travelling waves. Here, the noise intensity $D_{\rm s}$  is fixed at
14.0. Furthermore, we define several quantities of the traveling
waves. The output period $T_{\rm out}$ is defined as follows:
$T_{i}$ is the time interval between the $i$th wave and the $i+1$th
wave. $m$ waves are taken into account and the average value of them
is $T_{\rm out}$, where $T_{\rm out}=\frac{\sum\limits_{i=1}^{m}
T_{i}}{m-1}$. The three velocities (the normal, the positive and the
negative) of the traveling waves are defined as follows: when
traveling waves propagate forward, there is a mean propagation
velocity which is the positive velocity, and denoted by $V_{+}$. The
mean velocity of the backward propagating waves is likewise denoted
by $V_{-}$. In addition, $V$ is defined here as the average velocity
of the whole propagation of traveling wave. The output period
$T_{\rm out}$, the normal velocity $V$, the positive velocity
$V_{+}$ and the negative velocity $V_{-}$ are shown in
Fig.~\ref{fig:5}. The slopes of the red, blue and yellow lines in
Fig.~\ref{fig:5} are $1/V$, $1/V_{+}$ and $1/V_{-}$, respectively.

\begin{figure}
  \includegraphics[width=8.5cm]{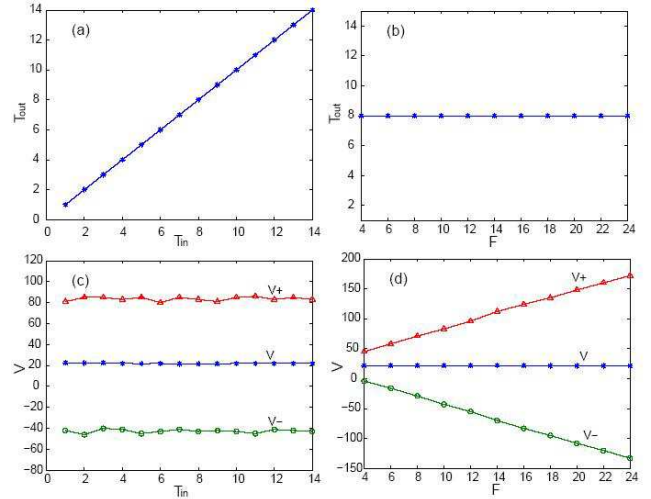}\\
  \caption{(Color online) The effect of periodic force on
  the property of traveling waves, $D_{\rm s}=14.0$.
  (a) The plot of the output period with respect to the input period, $F=10.0$.
  (b) The plot of the output period with respect to the intensity
   of periodic force, $T_{\rm in}=8.0$.
  (c) The plot of the three velocities of the traveling
  waves with respect to the input period, and $F=10.0$.
  (d) The plot of the three velocities of the traveling waves with respect to the intensity
  of periodic force, $T_{\rm in}=8.0$.}\label{fig:6}
\end{figure}
Based on the definition, we now focus our simulations on the
relationship between periodic force and the properties of travelling
waves. The results are shown in Fig.~\ref{fig:6}. Above all we
interpret how the periodic force $E(t)$ affects the output period
$T_{\rm out}$. First, we fix $F=10.0$ (the intensity of the periodic
force) and study how $T_{\rm out}$ changes with $T_{\rm in}$ (the
input period),as shown in Fig.~\ref{fig:6}(a). One observes that the
output period increases linearly with the input period. Moreover,
the output period is the same as the input period (the slope equals
to $1$). Second, we fix $T_{\rm in}=8.0$ and study the relation
between $T_{\rm out}$ and $F$, as shown in Fig.~\ref{fig:6}(b), from
which one observes that the output period is a constant and
independent of the intensity of the period forcing. So we can draw a
conclusion that the output period is the same as the input period
and is independent of the intensity of the periodic force, under a
fixed noise intensity.

Next we investigate the influence of the periodic force $E(t)$ on
the three velocities of the travelling waves (the normal $V$, the
positive $V_{+}$ and the negative $V_{-}$). First we fix
the intensity of the period forcing
 $F=10.0$  and see the changes of $V$, $V_{+}$, and
$V_{-}$ with the input period $T_{\rm in}$, respectively. The
results are shown in Fig.~\ref{fig:6}(c), from which we observe that
the normal, positive and negative velocities approximate constant
values, respectively, independent of the input period. Then we fix
$T_{\rm in}=8.0$ and study the changes of $V$, $V_{+}$ and $V_{-}$
with $F$ (see Fig.~\ref{fig:6}(d)), which shows that the normal
velocity is constant independent of the intensity of the periodic
forcing; the positive velocity increases with the intensity of
periodic force, while the negative velocity decreases with the
intensity of periodic force. So we can conclude that the normal
velocity is independent of the periodic force; the positive  the
negative velocities have the same trend with the intensity of
periodic forces and do not depend on the input period.

It is natural to presume from the above results that the three
velocities are interrelated. Through quantities of data, we obtain
the relationship among the normal, positive and negative velocities,
as follows
\begin{equation}\label{eq:2a}
2V=V_{+}+V_{-},
\end{equation}
that is,
\begin{equation}\label{eq:2b}
V^{*}=V_{+}-V=V-V_{-}.
\end{equation}

So the positive velocity is the normal velocity plus the $V^{*}$,
and the negative velocity is the normal velocity minus the $V^{*}$.

\section{conclusion and discussion}

In conclusion, noise and periodic force play a very important role
on the production, propagation, and stability of the traveling waves
in sub-excitable systems. Noise can induce traveling waves to
propagate stably. It can also support wave propagation in
sub-excitable~\cite{Zhou2002,PhysRevLett.82.3713,Kadar} media due to
a noise-induced
transition~\cite{PhysRevLett.87.078302,Garcia-Ojalvo,Francesc}. In
these studies, the media are static, and transport is governed by
diffusion. In many systems, the media are not static, but subject to
a motion, for example, when stirred by a flow, or by the oscillatory
electric field (with period). This occurs especially in chemical
reactions in a fluid environment. In this case, usually diffusive
transport dominates only at small spatial scales while mixing due to
the flow in much faster at large scale. In this paper, we
investigate the sub-excitable system using a simplified Oregonator
model, and the propagation of traveling waves in the presence of
both noise and periodic force. Depending on noise and the periodic
forcing we have observed different types of wave propagation (or its
disappearance). Moreover, the reversal phenomena is observed in this
system based on the numerical experiments in one-dimensional space.
we give qualitative explanations to how reversal phenomena appear,
which turns out to be the periodic force. The output period and
three velocities (the normal, the positive and the negative) of the
traveling waves are defined and their relationship with the periodic
forcing are also studied in sub-excitable system with a fixed
intensity of noise.

The periodic force can make traveling waves produced periodically
and reversal phenomena appear. The reversal phenomenon results from
the alternation of the positive and negative values of the periodic
forcing and noise. For the special case, we give the phase diagram
about the reversal phenomena with respect to $D_{\rm s}$--$F$
parameter space, from which one can see that the reversal waves
sensitively depend on the intensity of periodic forces and the noise
intensities for the fixing $T_{in}$. At last we research on the
effect of periodic force on the sub-excitable system with a fixed
noise intensity $D_{\rm s}=14.0$. Under thus a condition, the
influence of periodic forces on the three different kinds of
velocities are also acquired. The relation among the three
velocities is $2V=V_{+}+V_{-}$. In Ref.~\cite{Zhou2002}, the authors
studied sub-excitable medium of Belousov-Zhabotinsky (BZ) reaction
subjected to Gaussian white noise in experiments. They observed that
at an optimal level of noise the wave sources of excited traveling
waves become synchronous, as though there exists a long distance
spatial correlation.

\section*{Acknowledgment}
We thank Professor Qi Ouyang for enlightening discussion about this
paper. We are grateful for the meaningful suggestions of the two
anonymous referees.



\end{document}